\documentclass[ammsymb, twocolumn, prb, superscriptaddress, floats, showkeys, showpacs, a4paper]{revtex4-1}
  
\usepackage[T1]{fontenc}
\usepackage{bm}
\usepackage{graphicx}
\usepackage{natbib}
\usepackage{epstopdf}
\usepackage[euler]{textgreek}
\usepackage{hyperref}
\usepackage{color}

\usepackage[caption=false]{subfig}

\begin{document}
 
\title{Energy spectrum of confined positively charged excitons in single quantum dots}

\author{M. R. Molas}
\email{maciej.molas@gmail.com}
\affiliation{Faculty of Physics, University of Warsaw, ul. Pasteura 5, 02-093 Warszawa, Poland}
\affiliation{Laboratoire National des Champs Magn\'etiques Intenses, CNRS-UJF-UPS-INSA, 25, avenue des Martyrs, 38042 Grenoble, France} 
\author{A. W\'ojs}
\email{arkadiusz.wojs@pwr.edu.pl }
\affiliation{Department of Theoretical Physics, Wroclaw University of Technology, Wybrze\.ze Wyspia\'nskiego 27, 50-370 Wroc\l{}aw, Poland }
\author{A. A. L. Nicolet}
\affiliation{Laboratoire National des Champs Magn\'etiques Intenses, CNRS-UJF-UPS-INSA, 25, avenue des Martyrs, 38042 Grenoble, France} 
\author{A. Babi\'nski}
\affiliation{Faculty of Physics, University of Warsaw, ul. Pasteura 5, 02-093 Warszawa, Poland}
\author{M. Potemski}
\affiliation{Laboratoire National des Champs Magn\'etiques Intenses, CNRS-UJF-UPS-INSA, 25, avenue des Martyrs, 38042 Grenoble, France}

\date{\today}

\begin{abstract}

A theoretical model, which relates the binding energy of a positively charged exciton in a quantum dot with the confinement energy is presented. It is shown that the binding energy, defined as the energy difference between the corresponding charged and neutral complexes  confined on the same excitonic shell strongly depends on the shell index.  Moreover, it is shown that the ratio of the binding energy for positively charged excitons from the $p$- and $s$-shells of a dot depends mainly on the nearly perfect confinement in the dot, which is due to the "hidden symmetry" of the multi-electron-hole system.  We applied the theory to the excitons confined to a single GaAlAs/AlAs quantum dots. The relevant binding energy was determined using the micro-photoluminescence and micro-photoluminescence excitation magneto-spectroscopy. We show that within our theory, the confinement energy determined using the ratio of the binding energy corresponds well to the actual confinement energy of the investigated dot.

\end{abstract}

\maketitle

\section{Introduction \label{sec:Intro}}

Semiconductor quantum dots (QDs) provide a unique environment to study fundamental properties of strongly interacting charge carriers.\cite{grundmann,jacak,nano} The multitude of possible effects including direct and exchange  Coulomb interactions and the resulting configuration mixing makes the energy spectrum of the complexes a complicated function of the confining potential and the relevant interactions. Both factors which depend on the QD size, shape, and composition must be taken into account to reliably describe the energy spectrum of excitons.  Substantial efforts have been put to relate the morphology to experimentally addressable properties of carrier complexes confined in dots, $i.e.$, the related photoluminescence (PL) - which corresponds to emission or the photoluminescence excitation (PLE) - which corresponds to the absorption of light.\cite{luo,benny2012}  Among them a concept of "inverse engineering", which associates a specific order of emission lines due to particular excitonic complexes with a specific structure and composition of dots can be acknowledged.\cite{mlinar} Going beyond the approach we have recently shown that in natural InAs QDs the order results from a particular realization of the atomic species distribution.\cite{zielinski}

The quest to relate some general properties of the QDs confining potential to the experimentally addressable features of carrier complexes in the dots motivates also our study. Our theoretical analysis shows that the ratio of the binding energy (BE) of the positively charged excitonic complexes related to the $p$- and $s$-shells of a strongly confined QD depends mainly on the confinement energy in the QD. We applied the model to our experimental results confirming its validity in the GaAlAs/AlAs QDs formed in a GaAs/AlAs type-II bilayer. 

\section{Theoretical model \label{sec:theory}}

\begin{center}
\begin{figure}[t]
\centering
  \includegraphics[width=0.9\linewidth]{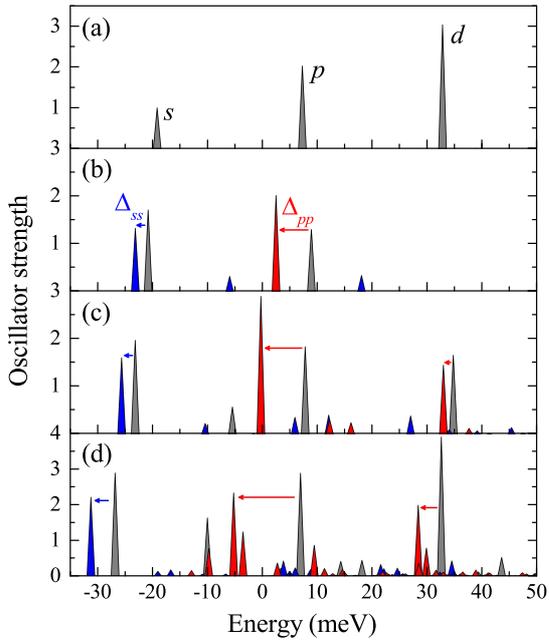}%
\caption{(color online) Absorption spectra within the $s$-, $p$-, and $d$-shell of a single QD parametrised by $\omega _e$= 20~meV and $\omega _h$= 4~meV in the absence of magnetic field while: (a) all inter-shell Coulomb scattering matrix elements are ignored; all Coulomb matrix elements within (b) 2 ($s$, $p$), (c) 3 ($s$, $p$, $d$), and (d) 8 ($s$, $p$, $d$ \dots) electron and hole shells are included. Grey and blue (red) peaks represent an emission related to the neutral exciton and the singlet(triplet)-spin state of the positively charged exciton, respectively. The horizontal blue and red arrows indicate a magnitude of binding energy at the $s$- and $p$-shell ($\Delta_{ss}$ and $\Delta_{pp}$), respectively. \label{fig:theory01}}
\end{figure}
\end{center}

We carried out the numerical calculations by exact diagonalization of model Coulomb Hamiltonians of the studied $eh$ complexes ($1e+1h$ for the exciton and $1e+2h$ for the positively charged exciton). For simplicity, we assumed that the QD confinement is circular, strictly two-dimensional ($i.e.$, of zero thickness) and laterally parabolic for both electrons and holes. Furthermore, we have assumed equal $e$ and $h$ oscillator length scales ($i.e.$, identical corresponding electron and hole orbitals), which relies on nearly perfect (and thus spatially equal) confinement of both kinds of carriers inside the quantum dot; within our simple model this is expressed by equal ratios of electron and hole effective masses and confinement frequencies ($m_e\omega _e$=$m_h\omega _h$).\cite{hawrylak} The neutral and positively charged exciton Hamiltonian matrices were expressed in the $eh$ configuration bases, and we have only included a small number of lowest electron and hole $s$-, $p$-, $d$-,\dots oscillator shells. For the present context of studying two particular optical transitions (between an exciton and vacuum and between a charged exciton and the hole in the $s$-shell), only the subspaces with vanishing total orbital angular momentum M=0 are important and have been considered. For the charged exciton, the total two-hole spin (singlet vs. triplet) has been resolved in the diagonalization procedure. Finally, for each obtained neutral or charged exciton eigenstate we have also computed the oscillator strength for the relevant optical transition mentioned above: X$\leftrightarrow$vacuum and X$^+$$\leftrightarrow$$h_s$. 

Theoretically determined absorption spectra for the neutral exciton as well as for the spin-singlet and the spin-triplet positively charged excitons are shown in Fig. \ref{fig:theory01}. The presented spectra were calculated with a particular choice of parameters: $\omega _e$=20~meV and $\omega _h$=4~meV ($i.e.$, $\omega _e$/$\omega _h$=$m_h$/$m_e$=5) and in the absence of magnetic field (B=0~T). All material parameters ($i.e.$, dielectric constant $\epsilon$=12.5 and effective mass of electron $m_e$=0.067~$m_0$) were taken appropriate for GaAs. The Planck constant ($\hbar$) is omitted here and in the following considerations for the sake of clarity. In the zero-order approximation only the electron-hole and the hole-hole intra-shell Coulomb interactions were considered with all inter-shell Coulomb scattering matrix elements neglected.  The results of such an approach are displayed in Fig. \ref{fig:theory01}(a).  The spectra are identical for both complexes, with few peaks corresponding to the promotion of electrons from the successive $s$-, $p$-, $d$-, . . . valence band shells to the corresponding conduction band shells. The peaks related to the shells are spaced roughly by the confining energy  $\omega _e$+$\omega _h$ with oscillator strengths equal to the shell degeneracies.

\begin{center}
\begin{figure}[t]
\centering
  \includegraphics[width=0.95\linewidth]{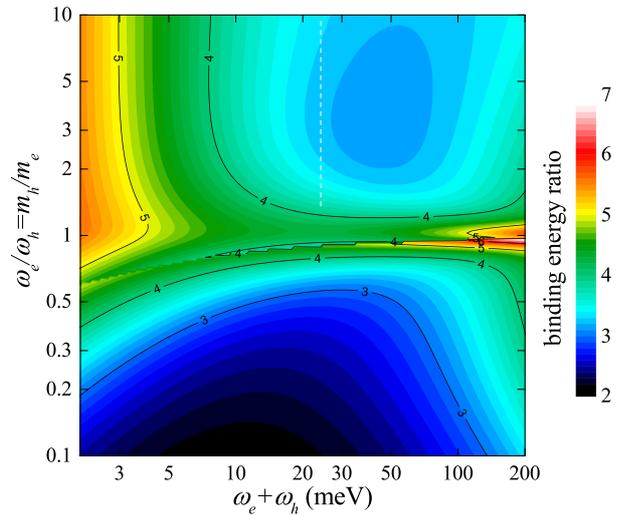}%
\caption{(color online) Color-coded contour plot map of the binding energy ratio R=$\Delta_{pp}/\Delta_{ss}$ in coordinates $\omega _e$+$\omega _h$ and $\omega _e$/$\omega _h$.  \label{fig:theory02}}
\end{figure}
\end{center}

Figs \ref{fig:theory01}(b)-(d) present the spectra obtained by including all Coulomb matrix elements within 2, 3, and 8 electron and hole shells. Clearly, already a simple two-shell model captures the essential behavior. Besides the known redistribution of excitonic oscillator strength from higher to lower peaks caused by inter-shell scattering, several points need be noted about the trions: (i) Low-energy absorption adding to an $s$-shell hole an $eh$ pair on the same, lowest $s$-shell only creates the spin-singlet charged exciton (blue peaks, including one marked "$ss$"). In contrast, higher energy absorption adding an $eh$ pair on different, higher shells predominantly creates the spin-triplet charged exciton (red peaks, including one marked "$pp$"). (ii) The energy difference $\Delta_{xx}$ between the corresponding neutral and charged exciton peaks ("$ss$" and "$pp$"), interpreted as binding energy of the an exciton in different $x$-shells to a hole in the lowest $s$-shell, is not monotonic in the shell index $x$=$s$, $p$, $d$, \ldots. While it does decrease with increasing $x$=$p$, $d$, \ldots, when absorption occurs to a different shell than that occupied by the initial electron ($x$>$s$), forming a spin-triplet charged exciton, it remains relatively low when absorption happens to the same shell ($x$=$s$), and a spin-singlet charged exciton is created. This exchange effect is a consequence of the (approximate) "hidden symmetry" in this system.\cite{wojs,raymond,hawrylak,bayernature} 

In order to investigate further the effect of quantum confinement on the optical spectra of the neutral and positively charged excitons we performed the previously shown analysis with 3 shells (compare Fig.\ref{fig:theory01}(c)) for a broad range of the relevant parameters. To this end we determined the ratio (R) of the BE of the positively charged trions (R=$\Delta_{pp}/\Delta_{ss}$). The results of our calculations are illustrated in Fig. \ref{fig:theory02} in the form of a color-coded contour plot) in intuitive $\omega _e$+$\omega _h$ and $\omega _e$/$\omega _h$ coordinates.  As it can be appreciated in Fig. \ref{fig:theory02} the ratio R varies rather little throughout most of the map. Moreover for the experimentally justified case of $\omega _e$/$\omega _h$>1 (corresponding to usual $m _e$/$m_h$<1) the contours go almost vertically in a broad range of the confinement energies.  Particularly, in the case of  $\omega _e$+$\omega _h\approx$24~meV, previously described in Fig. \ref{fig:theory01}, R=3.5 (see the white dashed line in Fig. \ref{fig:theory02}).

\section{Sample and Experimental Setups \label{sec:procedure}}
The active part of the structure used in our study was intentionally designed as a GaAs/AlAs type-II bilayer ($d_{\textrm {GaAs}}$ = 2.4 nm, $d_{\textrm {AlAs}}$ = 10 nm) embedded  between wide (100 nm) $\textrm{Ga}_{0.67}\textrm{Al}_{0.33}\textrm{As}$ barriers.\cite{truby,wysmolekE} Previous research showed that the bilayer is not perfect in the lateral directions: the Ga-rich inclusions, which can be seen as islands of $\textrm{Ga}_{1-x}\textrm{Al}_{x}\textrm{As}$  ($x$ < 0.33) replacing the original GaAs/AlAs bilayer, exist in this structure and possess all attributes of relatively strongly confined semiconductor QDs. These dots show remarkably low surface density, at the level of $10^5-10^6$ $\textrm{cm}^{-2}$. Their emission spectra are dispersed in a wide energy range, 1.56-1.68 eV,\cite{wysmolekappa,pietkaphd,molasnatural,martin,pietkaprb,molasphd,molasepl} due to the spread in the lateral extent of the confining potential.

Single dot measurements have been carried out at liquid helium temperature using a typical setup for the PL and PLE experiments. To detect the PL spectra, a tunable Ti:Sapphire laser was set at $\lambda$= 725 nm to assure the quasi-resonant excitation conditions, $i.e.$, to inject the $eh$ pairs directly into QDs.\cite{kazimierczuknon} The PLE signal was obtained from the measured variations in the intensity of the emitted PL with changes of the excitation energy.

The setup was dedicated to measurements in an external magnetic field in the Faraday configuration with the aid of a superconducting magnet producing field up to 9~T. The sample was located on top of an x-y-z piezo-stage in a bath cryostat with an optical access provided by a Y-shaped fiber equipped with a microscope objective (spot size around 1 \textmu m$^2$).\cite{adamphysE} The laser light was coupled to one branch of the Y-shaped fiber, focused by the microscope objective, and the signal was detected from the second branch of the fiber, by a 0.5 m-long monochromator equipped with a charge-couple-device camera.

\section{Experimental results and discussion \label{sec:experiment}}

\begin{center}
\begin{figure}[t]
\centering
  \includegraphics[width=0.8\linewidth]{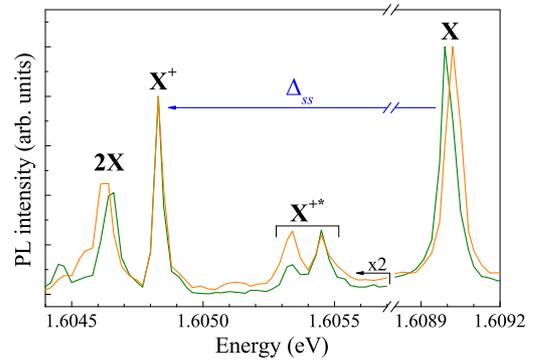}%
\caption{(color online) The PL spectra of a single GaAlAs QD recorded for two perpendicular linear polarizations oriented along the crystallographic directions $[110]$ (orange curve) and $[1\bar{1}0]$ (green curve), respectively. \label{fig:pl}}
\end{figure}
\end{center}

In order to experimentally verify the model and to find R for a particular QD, single-dot PL and PLE spectra must be collected and analysed. Moreover, as it is shown latter, the PLE measurements in magnetic field need to be performed in order to identify neural and positively charged excitons from the $p$-shell of a single QD.

The BE of the positively charged exciton at the ground s-shell ($\Delta_{ss}$) is defined as the energy difference between the emission lines attributed to the singlet state of the positively charged exciton (X$^+$) and to the neutral exciton (X). The results of polarization-sensitive PL measurements on the investigated QD are shown in Fig.  \ref{fig:pl}. It was found that both the X and the 2X emission lines split into two linearly-polarized components (the magnitude of splitting was equal to about 23 \textmu eV), which is characteristic of neutral excitonic complexes. The excitation-power dependent measurements confirmed that the former line corresponds to recombination of a neutral exciton (1$e_s$1$h_s$) while the latter results from the optical recombination of a neutral biexciton (2$e_s$2$h_s$). The fine structure splitting of the neutral exciton and biexciton is a consequence of the $eh$ exchange interaction in the dot characterized by anisotropic potential and it has been intensively studied in the literature.\cite{gammonprl,favero,flissikowski,bayerprb,karlsson}

The X$^+$ line did not split, which is characteristic of a charged exciton. The attribution of the line to the spin-singlet state of a singly, positively charged exciton was a topic of previous studies  (Refs \citenum{molascharged,molasphd}) and will not be discussed here.  The complex consists of an $s$-shell electron and two $s$-shell holes (1$e_s$2$h_s$).  There is no FSS of the X$^+$ line, because the $eh$ exchange interaction influences neither the initial (where the two holes form a closed shell) nor the final state (only one hole left).\cite{bayerprb} One must note that the presence of the emission lines due to different charge states of a QD (neural and positively charged) under quasi-resonant excitation is characteristic for spectra measured over some time (5 s in our case) and results from charge fluctuations in the structure.\cite{pietkaprb}

The BE of the positively charged exciton at the $s$-shell (of an $s$-shell exciton to the $s$-shell hole) determined from the analysis for the investigated QD equals $\Delta_{ss}\approx$4~meV.

For the sake of completeness yet another emission line (X$^{+*}$) present in the spectrum should be addressed. It is ascribed to the spin-triplet state of a singly, positively charged exciton and its fine structure results from the $eh$ and hole-hole exchange interaction (see Ref. \citenum{molascharged} for details).

\begin{center}
\begin{figure}[t]
\centering
  \includegraphics[width=0.8\linewidth]{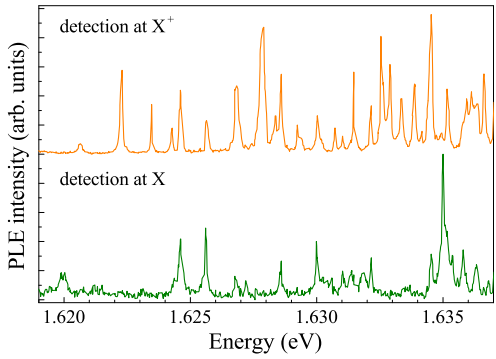}%
\caption{(color online) The PLE spectra detected on the (green curve) X and (orange curve) X$^+$ emission lines of a single GaAlAs QD. The spectra are normalized to the most intense peaks and shifted for clarity purpose.\label{fig:ple_0T}}
\end{figure}
\end{center}

More complicated is the determination of the BE of the positively charged exciton at the $p$-shell ($\Delta_{pp}$). The BE is defined as the difference between the energies of the triplet state of the positively charged exciton (1$e_p$1$h_p$1$h_s$) and to the neutral exciton in the excited state (1$e_p$1$h_p$). Both states do not usually recombine radiatively due to their fast relaxation to the spin-singlet of a positively charged exciton and the ground state of the neutral exciton. They should however contribute to the resonances in the PLE spectra of the charged and neutral excitons measured respectively at the X and X$^+$ emission lines. The PLE spectra are presented in Fig. \ref{fig:ple_0T}. 

\begin{figure*}[t]
\subfloat{}%
\centering
  \includegraphics[width=0.29\linewidth]{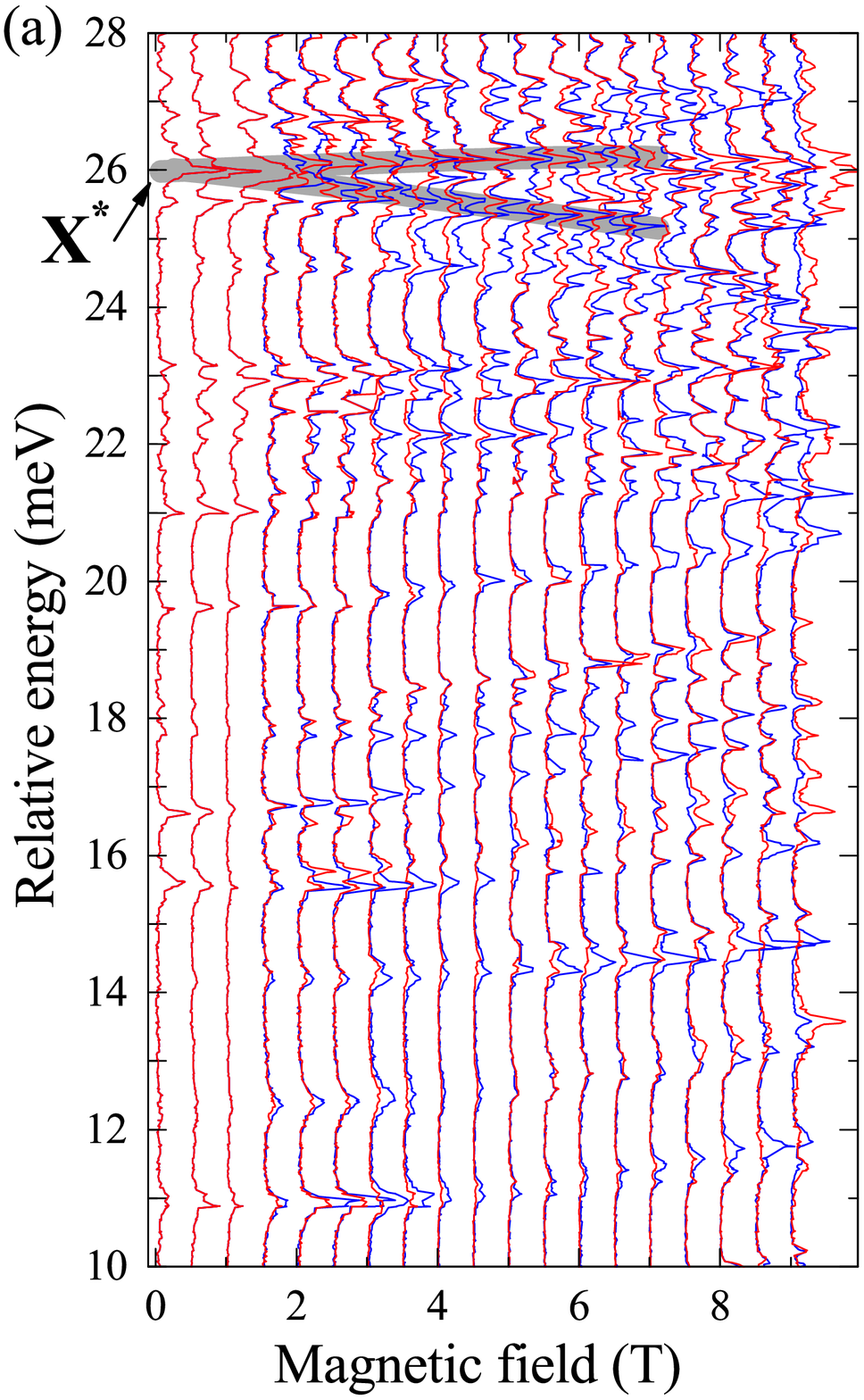}%
\subfloat{}%
\centering
  \includegraphics[width=0.29\linewidth]{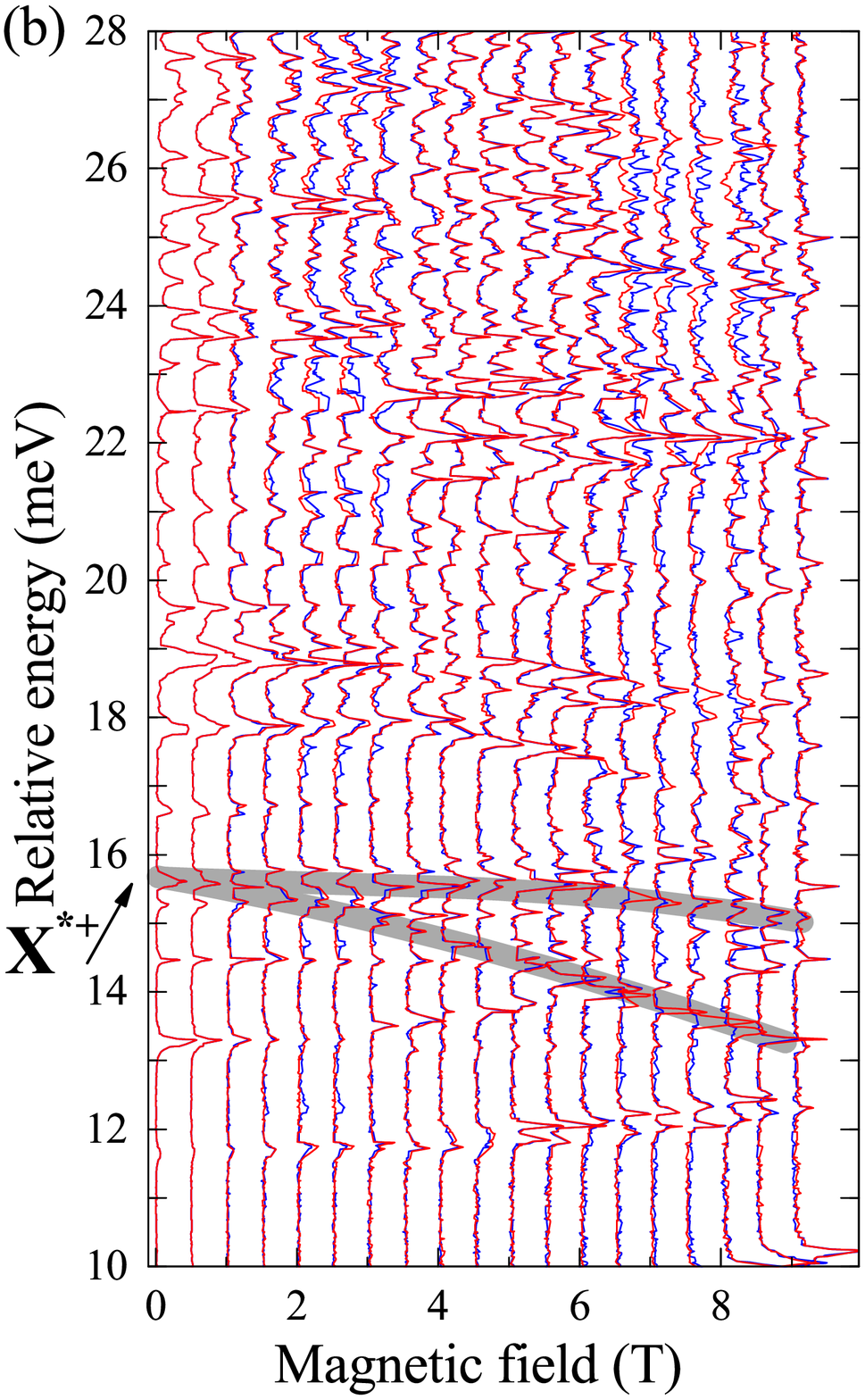}%
\caption{(color online) The magnetic-field evolution of PLE spectra detected on the (a) X and (b) X$^+$ emission lines. The red and blue curves indicate the $\sigma_{+}$- and $\sigma_{-}$-polarised components of the  lines, on which the PLE spectra were measured. The scale of the vertical axis is set by the energy relative to the X emission line at 0 T. The spectra are normalized to the most intense peaks and shifted for clarity purpose. Grey lines are a guide to the eye.\label{fig:pleXplus}}
\end{figure*}

There are several resonances in the spectra, which can not be directly attributed to the sought excited states of the neutral and positively charged excitons. It is not obvious at the moment what is the origin of the multitude of resonances. The symmetry breaking and the resulted mixing with higher energy bands can be a possible origin of the resonances,\cite{warming} however no solid explanation of the effect can be proposed. As a plethora of resonances does not facilitate their attribution to particular excitonic configurations, we investigated their evolution in magnetic field applied in the Faraday configuration. As it is known from previous experiments on the investigated dots, several shells ($s$-, $p$-, $d$-) are available to excitonic complexes confined in them. It is therefore reasonable to expect the observation of the evolution in magnetic field which is related to the $p$-shell of a dot.\cite{babinskiprb2006b} The single-particle electronic state of a particle confined in parabolic potential is expected to follow the Fock-Darwin structure.\cite{fock,darwin} Previous experiments on QDs shown that the characteristic evolution is also reproduced by the energy structure of  the excitonic states.\cite{raymondprl} In particular the energy of a complex relate to the excited states, which are of interest to us, is supposed to lower its energy with magnetic field (which results from Zeeman interaction of the orbital momentum of the complex with magnetic field). The results of the measurements are shown in Fig. \ref{fig:pleXplus}. As the energies of the positively charged exciton (X$^+$) and the neutral exciton (X) shift diamagnetically in magnetic field and split due to Zeeman interaction, the energy scales in both Fig. \ref{fig:pleXplus}(a) and Fig. \ref{fig:pleXplus}(b) are relative to the energies of corresponding ground state complexes. In the following we address both results.

The PLE spectra of the neutral exciton (see Fig. \ref{fig:pleXplus}(a)) can be divided into two energy ranges: above and below $\sim$21 meV. In the low-energy range, all resonant peaks show the same type of the magnetic field evolution. Their energies increase with magnetic field which is characteristic diamagnetic behaviour of carriers and complexes with zero-orbital momentum. Such an evolution relates the resonances to the $s$-shell of the investigated dot.\cite{babinskiprb2006b} We can observed them at the relative energy higher than approx. 10 meV and the energy difference between them is only around 1-2 meV. The observed resonance also split in magnetic field due to Zeeman spin- interaction. A complicated picture of the splitting of particular resonances observed in particular polarizations of the ground state makes the full analysis of the effect quite complicated a task. However, a similar effect of magnetic field on all those resonances can be observed we provisionally relate them to resonances occurred between the excited hole levels, $e.g.$, the $p$-, $d$-,$\ldots$ shells, and the ground electron level, the $s$-shell.\cite{benny2011,molasple,molasphd} Such an attribution would explain the diamagnetic shift of the resonances. At high energies, the observed pattern of resonant peaks becomes extremely complex and the peaks can be hardly followed with the magnetic field. In our opinion some additional transitions involving phonon replicas can be responsible for resonances in this energy range. Nevertheless there are resonances observed in that energy range which lower their energies with magnetic field. The lowest-energy resonance of that behaviour can be appreciated following the structure, which originates around $\sim$26 meV at B=0~T.  We ascribed this peak to the excited state of the neutral exciton, labelled X$^*$ (1$e_p$1$h_p$), involving the absorption of an electron from the $p$-shell level in the valence band to the $p$-shell level in the conduction band.

In the case of magnetic-field evolution of PLE spectra of the charged exciton, presented in Fig. \ref{fig:pleXplus}(b), there are also two types of resonances: the $s$- and $p$-shell like. The observation of the $s$-shell-like magnetic-field dependence might also be ascribed to the transitions between the excited hole levels, $e.g.$, the $p$-, $d$-,$\ldots$ shells, and the ground electron level, the $s$-shell, however, in the presence of an additional hole.\cite{benny2011,molasple,molasphd} Moreover, the peak, appeared around $\sim$15.5 meV at B=0~T, lowers its energy with the magnetic field. As a consequence this peak has been attributed to the absorption process of excited state of the positively charged exciton, labelled X$^{*+}$ (1$e_p$1$h_p$1$h_s$), which involve absorption process of an electron from the $p$-shell level in the valence band to the $p$-shell level in the conduction band in presence of an extra hole on the $s$-shell (see Refs. \citenum{molasphd,molasX+} for details).

For the $p$-shell, we obtained that the BE of an $p$-shell exciton to the $s$-shell hole is about $\Delta_{pp}\approx$10 meV for the studied dot. Consequently, it was found that BE of the positively charged exciton, defined as the energy difference between the charged and neutral complex occupying the $s$- and $p$-shell levels, increases more than two times from the $p$-shell to the $s$-shell complex (the $\Delta_{pp}/\Delta_{ss}$ ratio equals about 2.5). 

Our experimentally obtained value of the R ratio, $\approx$2.5, qualitatively agrees with the calculations of the presented simplified model. Based on the description of the GaAlAs/AlAs dots, discussed in Refs \citenum{pietkaphd,molasphd}, we assumed that the $\omega _e$+$\omega _h$ value contains in the range 10 - 20~meV, while the $\omega _e$/$\omega _h$ parameter is bigger than about 3. This situates these QDs in central-upper part of Fig. Fig.~\ref{fig:theory02}, where the BE ratio changes from 3 to 4, which is very close to the experimental value.

This result confirms the validity of the proposed model based on some very general assumptions related to basic properties of electron-hole complexes in QDs. 

\section{Conclusions \label{sec:summary}}

We proposed a theoretical model, which described the "shell" effect on the binding energy (BE) of the positively charged excitons. Our calculations showed that the BE ratio remarkably remains almost constant over a broad range of applied parameters.  We performed PL and PLE  excitation measurements on the GaAs/AlGaAs QDs to verify the validity of the model. We showed that application of magnetic field was necessary to identify the excited states of both the neutral and positively charged excitonic states. Based on experimental values of the BE of the positively charged exciton we obtained a qualitative agreement between the experimental results and the predictions of our theory. Our results confirm the relevance of the applied model to the description of electron-hole complexes in QDs.

\section{Acknowledgements}
The work has been supported by the Foundation for Polish Science International PhD Projects Programme co-financed by the EU European Regional Development Fund. M.R.M. kindly acknowledges the National Science Center (NCN) Grant No. DEC-2013/08/T/ST3/00665 and DEC-2013/09/N/ST3\linebreak[4]/04237 for financial support for his PhD. A.W. acknowledges support from the NCN under Grant No.~2014/14/A/ST3/00654.

\bibliographystyle{apsrev4-1}
\bibliography{biblioBinding}

\begin{thebibliography}{36}%
\makeatletter
\providecommand \@ifxundefined [1]{%
 \@ifx{#1\undefined}
}%
\providecommand \@ifnum [1]{%
 \ifnum #1\expandafter \@firstoftwo
 \else \expandafter \@secondoftwo
 \fi
}%
\providecommand \@ifx [1]{%
 \ifx #1\expandafter \@firstoftwo
 \else \expandafter \@secondoftwo
 \fi
}%
\providecommand \natexlab [1]{#1}%
\providecommand \enquote  [1]{``#1''}%
\providecommand \bibnamefont  [1]{#1}%
\providecommand \bibfnamefont [1]{#1}%
\providecommand \citenamefont [1]{#1}%
\providecommand \href@noop [0]{\@secondoftwo}%
\providecommand \href [0]{\begingroup \@sanitize@url \@href}%
\providecommand \@href[1]{\@@startlink{#1}\@@href}%
\providecommand \@@href[1]{\endgroup#1\@@endlink}%
\providecommand \@sanitize@url [0]{\catcode `\\12\catcode `\$12\catcode
  `\&12\catcode `\#12\catcode `\^12\catcode `\_12\catcode `\%12\relax}%
\providecommand \@@startlink[1]{}%
\providecommand \@@endlink[0]{}%
\providecommand \url  [0]{\begingroup\@sanitize@url \@url }%
\providecommand \@url [1]{\endgroup\@href {#1}{\urlprefix }}%
\providecommand \urlprefix  [0]{URL }%
\providecommand \Eprint [0]{\href }%
\providecommand \doibase [0]{http://dx.doi.org/}%
\providecommand \selectlanguage [0]{\@gobble}%
\providecommand \bibinfo  [0]{\@secondoftwo}%
\providecommand \bibfield  [0]{\@secondoftwo}%
\providecommand \translation [1]{[#1]}%
\providecommand \BibitemOpen [0]{}%
\providecommand \bibitemStop [0]{}%
\providecommand \bibitemNoStop [0]{.\EOS\space}%
\providecommand \EOS [0]{\spacefactor3000\relax}%
\providecommand \BibitemShut  [1]{\csname bibitem#1\endcsname}%
\let\auto@bib@innerbib\@empty
\bibitem [{\citenamefont {Grundmann}\ \emph {et~al.}(1998)\citenamefont
  {Grundmann}, \citenamefont {Bimberg},\ and\ \citenamefont
  {Ledentsov}}]{grundmann}%
  \BibitemOpen
  \bibfield  {author} {\bibinfo {author} {\bibfnamefont {M.}~\bibnamefont
  {Grundmann}}, \bibinfo {author} {\bibfnamefont {D.}~\bibnamefont {Bimberg}},
  \ and\ \bibinfo {author} {\bibfnamefont {N.~N.}\ \bibnamefont {Ledentsov}},\
  }\href@noop {} {\emph {\bibinfo {title} {Quantum Dot Heterostructures}}}\
  (\bibinfo  {publisher} {John Wiley $\&$ Sons Ltd.},\ \bibinfo {address} {New
  York},\ \bibinfo {year} {1998})\BibitemShut {NoStop}%
\bibitem [{\citenamefont {Jacak}\ \emph {et~al.}(1998)\citenamefont {Jacak},
  \citenamefont {Hawrylak},\ and\ \citenamefont {Wojs}}]{jacak}%
  \BibitemOpen
  \bibfield  {author} {\bibinfo {author} {\bibfnamefont {L.}~\bibnamefont
  {Jacak}}, \bibinfo {author} {\bibfnamefont {P.}~\bibnamefont {Hawrylak}}, \
  and\ \bibinfo {author} {\bibfnamefont {A.}~\bibnamefont {Wojs}},\ }\href@noop
  {} {\emph {\bibinfo {title} {Quantum Dots}}}\ (\bibinfo  {publisher}
  {Springer-Verlag},\ \bibinfo {address} {Berlin},\ \bibinfo {year}
  {1998})\BibitemShut {NoStop}%
\bibitem [{\citenamefont {Grundmann}(2002)}]{nano}%
  \BibitemOpen
  \bibinfo {editor} {\bibfnamefont {M.}~\bibnamefont {Grundmann}},\ ed.,\ \href
  {\doibase 10.1007/978-3-642-56149-8} {\emph {\bibinfo {title}
  {Nano-Optoelectronics Concepts, Physics, and Devices}}}\ (\bibinfo
  {publisher} {Springer-Verlag},\ \bibinfo {address} {Berlin},\ \bibinfo {year}
  {2002})\BibitemShut {NoStop}%
\bibitem [{\citenamefont {Luo}\ and\ \citenamefont {Zunger}(2011)}]{luo}%
  \BibitemOpen
  \bibfield  {author} {\bibinfo {author} {\bibfnamefont {J.-W.}\ \bibnamefont
  {Luo}}\ and\ \bibinfo {author} {\bibfnamefont {A.}~\bibnamefont {Zunger}},\
  }\href {\doibase 10.1103/PhysRevB.84.235317} {\bibfield  {journal} {\bibinfo
  {journal} {Phys. Rev. B}\ }\textbf {\bibinfo {volume} {84}},\ \bibinfo
  {pages} {235317} (\bibinfo {year} {2011})}\BibitemShut {NoStop}%
\bibitem [{\citenamefont {Benny}\ \emph {et~al.}(2012)\citenamefont {Benny},
  \citenamefont {Kodriano}, \citenamefont {Poem}, \citenamefont {Gershoni},
  \citenamefont {Truong},\ and\ \citenamefont {Petroff}}]{benny2012}%
  \BibitemOpen
  \bibfield  {author} {\bibinfo {author} {\bibfnamefont {Y.}~\bibnamefont
  {Benny}}, \bibinfo {author} {\bibfnamefont {Y.}~\bibnamefont {Kodriano}},
  \bibinfo {author} {\bibfnamefont {E.}~\bibnamefont {Poem}}, \bibinfo {author}
  {\bibfnamefont {D.}~\bibnamefont {Gershoni}}, \bibinfo {author}
  {\bibfnamefont {T.~A.}\ \bibnamefont {Truong}}, \ and\ \bibinfo {author}
  {\bibfnamefont {P.~M.}\ \bibnamefont {Petroff}},\ }\href {\doibase
  10.1103/PhysRevB.86.085306} {\bibfield  {journal} {\bibinfo  {journal} {Phys.
  Rev. B}\ }\textbf {\bibinfo {volume} {86}},\ \bibinfo {pages} {085306}
  (\bibinfo {year} {2012})}\BibitemShut {NoStop}%
\bibitem [{\citenamefont {Mlinar}\ \emph {et~al.}(2009)\citenamefont {Mlinar},
  \citenamefont {Bozkurt}, \citenamefont {Ulloa}, \citenamefont {Ediger},
  \citenamefont {Bester}, \citenamefont {Badolato}, \citenamefont {Koenraad},
  \citenamefont {Warburton},\ and\ \citenamefont {Zunger}}]{mlinar}%
  \BibitemOpen
  \bibfield  {author} {\bibinfo {author} {\bibfnamefont {V.}~\bibnamefont
  {Mlinar}}, \bibinfo {author} {\bibfnamefont {M.}~\bibnamefont {Bozkurt}},
  \bibinfo {author} {\bibfnamefont {J.~M.}\ \bibnamefont {Ulloa}}, \bibinfo
  {author} {\bibfnamefont {M.}~\bibnamefont {Ediger}}, \bibinfo {author}
  {\bibfnamefont {G.}~\bibnamefont {Bester}}, \bibinfo {author} {\bibfnamefont
  {A.}~\bibnamefont {Badolato}}, \bibinfo {author} {\bibfnamefont {P.~M.}\
  \bibnamefont {Koenraad}}, \bibinfo {author} {\bibfnamefont {R.~J.}\
  \bibnamefont {Warburton}}, \ and\ \bibinfo {author} {\bibfnamefont
  {A.}~\bibnamefont {Zunger}},\ }\href {\doibase 10.1103/PhysRevB.80.165425}
  {\bibfield  {journal} {\bibinfo  {journal} {Phys. Rev. B}\ }\textbf {\bibinfo
  {volume} {80}},\ \bibinfo {pages} {165425} (\bibinfo {year}
  {2009})}\BibitemShut {NoStop}%
\bibitem [{\citenamefont {Zieli\ifmmode~\acute{n}\else \'{n}\fi{}ski}\ \emph
  {et~al.}(2015)\citenamefont {Zieli\ifmmode~\acute{n}\else \'{n}\fi{}ski},
  \citenamefont {Go\l{}asa}, \citenamefont {Molas}, \citenamefont {Goryca},
  \citenamefont {Kazimierczuk}, \citenamefont {Smole\ifmmode~\acute{n}\else
  \'{n}\fi{}ski}, \citenamefont {Golnik}, \citenamefont {Kossacki},
  \citenamefont {Nicolet}, \citenamefont {Potemski}, \citenamefont
  {Wasilewski},\ and\ \citenamefont {Babi\ifmmode~\acute{n}\else
  \'{n}\fi{}ski}}]{zielinski}%
  \BibitemOpen
  \bibfield  {author} {\bibinfo {author} {\bibfnamefont {M.}~\bibnamefont
  {Zieli\ifmmode~\acute{n}\else \'{n}\fi{}ski}}, \bibinfo {author}
  {\bibfnamefont {K.}~\bibnamefont {Go\l{}asa}}, \bibinfo {author}
  {\bibfnamefont {M.~R.}\ \bibnamefont {Molas}}, \bibinfo {author}
  {\bibfnamefont {M.}~\bibnamefont {Goryca}}, \bibinfo {author} {\bibfnamefont
  {T.}~\bibnamefont {Kazimierczuk}}, \bibinfo {author} {\bibfnamefont
  {T.}~\bibnamefont {Smole\ifmmode~\acute{n}\else \'{n}\fi{}ski}}, \bibinfo
  {author} {\bibfnamefont {A.}~\bibnamefont {Golnik}}, \bibinfo {author}
  {\bibfnamefont {P.}~\bibnamefont {Kossacki}}, \bibinfo {author}
  {\bibfnamefont {A.~A.~L.}\ \bibnamefont {Nicolet}}, \bibinfo {author}
  {\bibfnamefont {M.}~\bibnamefont {Potemski}}, \bibinfo {author}
  {\bibfnamefont {Z.~R.}\ \bibnamefont {Wasilewski}}, \ and\ \bibinfo {author}
  {\bibfnamefont {A.}~\bibnamefont {Babi\ifmmode~\acute{n}\else
  \'{n}\fi{}ski}},\ }\href {\doibase 10.1103/PhysRevB.91.085303} {\bibfield
  {journal} {\bibinfo  {journal} {Phys. Rev. B}\ }\textbf {\bibinfo {volume}
  {91}},\ \bibinfo {pages} {085303} (\bibinfo {year} {2015})}\BibitemShut
  {NoStop}%
\bibitem [{\citenamefont {Hawrylak}(1999)}]{hawrylak}%
  \BibitemOpen
  \bibfield  {author} {\bibinfo {author} {\bibfnamefont {P.}~\bibnamefont
  {Hawrylak}},\ }\href {\doibase 10.1103/PhysRevB.60.5597} {\bibfield
  {journal} {\bibinfo  {journal} {Phys. Rev. B}\ }\textbf {\bibinfo {volume}
  {60}},\ \bibinfo {pages} {5597} (\bibinfo {year} {1999})}\BibitemShut
  {NoStop}%
\bibitem [{\citenamefont {W{\'o}js}\ and\ \citenamefont
  {Hawrylak}(1996)}]{wojs}%
  \BibitemOpen
  \bibfield  {author} {\bibinfo {author} {\bibfnamefont {A.}~\bibnamefont
  {W{\'o}js}}\ and\ \bibinfo {author} {\bibfnamefont {P.}~\bibnamefont
  {Hawrylak}},\ }\href {\doibase Exciton-exciton interactions in highly excited
  quantum dots in a magnetic field} {\bibfield  {journal} {\bibinfo  {journal}
  {Solid State Commun.}\ }\textbf {\bibinfo {volume} {100}},\ \bibinfo {pages}
  {487} (\bibinfo {year} {1996})}\BibitemShut {NoStop}%
\bibitem [{\citenamefont {Raymond}\ \emph {et~al.}(1996)\citenamefont
  {Raymond}, \citenamefont {Fafard}, \citenamefont {Poole}, \citenamefont
  {W{\'o}js}, \citenamefont {Hawrylak}, \citenamefont {Charbonneau},
  \citenamefont {Leonard}, \citenamefont {Leon}, \citenamefont {Petroff},\ and\
  \citenamefont {Merz}}]{raymond}%
  \BibitemOpen
  \bibfield  {author} {\bibinfo {author} {\bibfnamefont {S.}~\bibnamefont
  {Raymond}}, \bibinfo {author} {\bibfnamefont {S.}~\bibnamefont {Fafard}},
  \bibinfo {author} {\bibfnamefont {P.}~\bibnamefont {Poole}}, \bibinfo
  {author} {\bibfnamefont {A.}~\bibnamefont {W{\'o}js}}, \bibinfo {author}
  {\bibfnamefont {P.}~\bibnamefont {Hawrylak}}, \bibinfo {author}
  {\bibfnamefont {S.}~\bibnamefont {Charbonneau}}, \bibinfo {author}
  {\bibfnamefont {D.}~\bibnamefont {Leonard}}, \bibinfo {author} {\bibfnamefont
  {R.}~\bibnamefont {Leon}}, \bibinfo {author} {\bibfnamefont {P.~M.}\
  \bibnamefont {Petroff}}, \ and\ \bibinfo {author} {\bibfnamefont {J.~L.}\
  \bibnamefont {Merz}},\ }\href {\doibase 10.1103/PhysRevB.54.11548} {\bibfield
   {journal} {\bibinfo  {journal} {Phys. Rev. B}\ }\textbf {\bibinfo {volume}
  {54}},\ \bibinfo {pages} {11548} (\bibinfo {year} {1996})}\BibitemShut
  {NoStop}%
\bibitem [{\citenamefont {Bayer}\ \emph {et~al.}(2000)\citenamefont {Bayer},
  \citenamefont {Stern}, \citenamefont {Hawrylak}, \citenamefont {Fafard},\
  and\ \citenamefont {Forchel}}]{bayernature}%
  \BibitemOpen
  \bibfield  {author} {\bibinfo {author} {\bibfnamefont {M.}~\bibnamefont
  {Bayer}}, \bibinfo {author} {\bibfnamefont {O.}~\bibnamefont {Stern}},
  \bibinfo {author} {\bibfnamefont {P.}~\bibnamefont {Hawrylak}}, \bibinfo
  {author} {\bibfnamefont {S.}~\bibnamefont {Fafard}}, \ and\ \bibinfo {author}
  {\bibfnamefont {A.}~\bibnamefont {Forchel}},\ }\href {\doibase
  10.1038/35016020} {\bibfield  {journal} {\bibinfo  {journal} {Nature
  (London)}\ }\textbf {\bibinfo {volume} {405}},\ \bibinfo {pages} {923}
  (\bibinfo {year} {2000})}\BibitemShut {NoStop}%
\bibitem [{\citenamefont {Tr\"{u}by}\ \emph {et~al.}(1996)\citenamefont
  {Tr\"{u}by}, \citenamefont {Potemski},\ and\ \citenamefont {Planel}}]{truby}%
  \BibitemOpen
  \bibfield  {author} {\bibinfo {author} {\bibfnamefont {A.}~\bibnamefont
  {Tr\"{u}by}}, \bibinfo {author} {\bibfnamefont {M.}~\bibnamefont {Potemski}},
  \ and\ \bibinfo {author} {\bibfnamefont {R.}~\bibnamefont {Planel}},\ }\href
  {\doibase 10.1016/0038-1101(95)00233-2} {\bibfield  {journal} {\bibinfo
  {journal} {Solid-State Electr.}\ }\textbf {\bibinfo {volume} {40}},\ \bibinfo
  {pages} {139} (\bibinfo {year} {1996})}\BibitemShut {NoStop}%
\bibitem [{\citenamefont {Wysmo\l{}ek}\ \emph {et~al.}(2002)\citenamefont
  {Wysmo\l{}ek}, \citenamefont {Potemski},\ and\ \citenamefont
  {Thierry-Mieg}}]{wysmolekE}%
  \BibitemOpen
  \bibfield  {author} {\bibinfo {author} {\bibfnamefont {A.}~\bibnamefont
  {Wysmo\l{}ek}}, \bibinfo {author} {\bibfnamefont {M.}~\bibnamefont
  {Potemski}}, \ and\ \bibinfo {author} {\bibfnamefont {V.}~\bibnamefont
  {Thierry-Mieg}},\ }\href {\doibase 10.1016/S1386-9477(01)00444-1} {\bibfield
  {journal} {\bibinfo  {journal} {Physica E}\ }\textbf {\bibinfo {volume}
  {12}},\ \bibinfo {pages} {876} (\bibinfo {year} {2002})}\BibitemShut
  {NoStop}%
\bibitem [{\citenamefont {Wysmo\l{}ek}\ \emph {et~al.}(2004)\citenamefont
  {Wysmo\l{}ek}, \citenamefont {Chwalisz}, \citenamefont {Potemski},
  \citenamefont {St\k{e}pniewski}, \citenamefont {Babi\'nski},\ and\
  \citenamefont {Raymond}}]{wysmolekappa}%
  \BibitemOpen
  \bibfield  {author} {\bibinfo {author} {\bibfnamefont {A.}~\bibnamefont
  {Wysmo\l{}ek}}, \bibinfo {author} {\bibfnamefont {B.}~\bibnamefont
  {Chwalisz}}, \bibinfo {author} {\bibfnamefont {M.}~\bibnamefont {Potemski}},
  \bibinfo {author} {\bibfnamefont {R.}~\bibnamefont {St\k{e}pniewski}},
  \bibinfo {author} {\bibfnamefont {A.}~\bibnamefont {Babi\'nski}}, \ and\
  \bibinfo {author} {\bibfnamefont {S.}~\bibnamefont {Raymond}},\ }\href@noop
  {} {\bibfield  {journal} {\bibinfo  {journal} {Acta Phys. Pol. A}\ }\textbf
  {\bibinfo {volume} {106}},\ \bibinfo {pages} {367} (\bibinfo {year}
  {2004})}\BibitemShut {NoStop}%
\bibitem [{\citenamefont {Pi\k{e}tka}(2007)}]{pietkaphd}%
  \BibitemOpen
  \bibfield  {author} {\bibinfo {author} {\bibfnamefont {B.}~\bibnamefont
  {Pi\k{e}tka}},\ }\href@noop {} {\bibfield  {journal} {\bibinfo  {journal}
  {Phd thesis, Joseph Fourier University, Grenoble I, and University of Warsaw,
  Warsaw}\ } (\bibinfo {year} {2007})}\BibitemShut {NoStop}%
\bibitem [{\citenamefont {Molas}\ \emph {et~al.}(2012)\citenamefont {Molas},
  \citenamefont {Go{\l{}}asa}, \citenamefont {Pi\k{e}tka}, \citenamefont
  {Potemski},\ and\ \citenamefont {Babi{\'n}ski}}]{molasnatural}%
  \BibitemOpen
  \bibfield  {author} {\bibinfo {author} {\bibfnamefont {M.}~\bibnamefont
  {Molas}}, \bibinfo {author} {\bibfnamefont {K.}~\bibnamefont {Go{\l{}}asa}},
  \bibinfo {author} {\bibfnamefont {B.}~\bibnamefont {Pi\k{e}tka}}, \bibinfo
  {author} {\bibfnamefont {M.}~\bibnamefont {Potemski}}, \ and\ \bibinfo
  {author} {\bibfnamefont {A.}~\bibnamefont {Babi{\'n}ski}},\ }\href@noop {}
  {\bibfield  {journal} {\bibinfo  {journal} {Acta Phys. Pol. A}\ }\textbf
  {\bibinfo {volume} {122}},\ \bibinfo {pages} {988} (\bibinfo {year}
  {2012})}\BibitemShut {NoStop}%
\bibitem [{\citenamefont {Mart\'in}\ \emph {et~al.}(2012)\citenamefont
  {Mart\'in}, \citenamefont {Ant\'on}, \citenamefont {Vi{\~n}a}, \citenamefont
  {Pi\k{e}tka},\ and\ \citenamefont {Potemski}}]{martin}%
  \BibitemOpen
  \bibfield  {author} {\bibinfo {author} {\bibfnamefont {M.~D.}\ \bibnamefont
  {Mart\'in}}, \bibinfo {author} {\bibfnamefont {C.}~\bibnamefont {Ant\'on}},
  \bibinfo {author} {\bibfnamefont {L.}~\bibnamefont {Vi{\~n}a}}, \bibinfo
  {author} {\bibfnamefont {B.}~\bibnamefont {Pi\k{e}tka}}, \ and\ \bibinfo
  {author} {\bibfnamefont {M.}~\bibnamefont {Potemski}},\ }\href {\doibase
  10.1088/1742-6596/210/1/012014} {\bibfield  {journal} {\bibinfo  {journal}
  {EPL}\ }\textbf {\bibinfo {volume} {100}},\ \bibinfo {pages} {67006}
  (\bibinfo {year} {2012})}\BibitemShut {NoStop}%
\bibitem [{\citenamefont {Pi\k{e}tka}\ \emph {et~al.}(2013)\citenamefont
  {Pi\k{e}tka}, \citenamefont {Suffczy\'{n}ski}, \citenamefont {Goryca},
  \citenamefont {Kazimierczuk}, \citenamefont {Golnik}, \citenamefont
  {Kossacki}, \citenamefont {Wysmolek}, \citenamefont {Gaj}, \citenamefont
  {St\k{e}pniewski},\ and\ \citenamefont {Potemski}}]{pietkaprb}%
  \BibitemOpen
  \bibfield  {author} {\bibinfo {author} {\bibfnamefont {B.}~\bibnamefont
  {Pi\k{e}tka}}, \bibinfo {author} {\bibfnamefont {J.}~\bibnamefont
  {Suffczy\'{n}ski}}, \bibinfo {author} {\bibfnamefont {M.}~\bibnamefont
  {Goryca}}, \bibinfo {author} {\bibfnamefont {T.}~\bibnamefont
  {Kazimierczuk}}, \bibinfo {author} {\bibfnamefont {A.}~\bibnamefont
  {Golnik}}, \bibinfo {author} {\bibfnamefont {P.}~\bibnamefont {Kossacki}},
  \bibinfo {author} {\bibfnamefont {A.}~\bibnamefont {Wysmolek}}, \bibinfo
  {author} {\bibfnamefont {J.~A.}\ \bibnamefont {Gaj}}, \bibinfo {author}
  {\bibfnamefont {R.}~\bibnamefont {St\k{e}pniewski}}, \ and\ \bibinfo {author}
  {\bibfnamefont {M.}~\bibnamefont {Potemski}},\ }\href {\doibase
  10.1103/PhysRevB.87.035310} {\bibfield  {journal} {\bibinfo  {journal} {Phys.
  Rev. B}\ }\textbf {\bibinfo {volume} {87}},\ \bibinfo {pages} {035310}
  (\bibinfo {year} {2013})}\BibitemShut {NoStop}%
\bibitem [{\citenamefont {Molas}(2014)}]{molasphd}%
  \BibitemOpen
  \bibfield  {author} {\bibinfo {author} {\bibfnamefont {M.}~\bibnamefont
  {Molas}},\ }\href@noop {} {\bibfield  {journal} {\bibinfo  {journal} {Phd
  thesis, Joseph Fourier University, Grenoble I, and University of Warsaw,
  Warsaw}\ } (\bibinfo {year} {2014})}\BibitemShut {NoStop}%
\bibitem [{\citenamefont {Molas}\ \emph {et~al.}(2016)\citenamefont {Molas},
  \citenamefont {Nicolet}, \citenamefont {Babi\'nski},\ and\ \citenamefont
  {Potemski}}]{molasepl}%
  \BibitemOpen
  \bibfield  {author} {\bibinfo {author} {\bibfnamefont {M.~R.}\ \bibnamefont
  {Molas}}, \bibinfo {author} {\bibfnamefont {A.~A.~L.}\ \bibnamefont
  {Nicolet}}, \bibinfo {author} {\bibfnamefont {A.}~\bibnamefont {Babi\'nski}},
  \ and\ \bibinfo {author} {\bibfnamefont {M.}~\bibnamefont {Potemski}},\
  }\href {\doibase 10.1209/0295-5075/113/17004} {\bibfield  {journal} {\bibinfo
   {journal} {EPL}\ }\textbf {\bibinfo {volume} {113}},\ \bibinfo {pages}
  {17004} (\bibinfo {year} {2016})}\BibitemShut {NoStop}%
\bibitem [{\citenamefont {Kazimierczuk}\ \emph {et~al.}(2009)\citenamefont
  {Kazimierczuk}, \citenamefont {Suffczy\ifmmode~\acute{n}\else \'{n}\fi{}ski},
  \citenamefont {Golnik}, \citenamefont {Gaj}, \citenamefont {Kossacki},\ and\
  \citenamefont {Wojnar}}]{kazimierczuknon}%
  \BibitemOpen
  \bibfield  {author} {\bibinfo {author} {\bibfnamefont {T.}~\bibnamefont
  {Kazimierczuk}}, \bibinfo {author} {\bibfnamefont {J.}~\bibnamefont
  {Suffczy\ifmmode~\acute{n}\else \'{n}\fi{}ski}}, \bibinfo {author}
  {\bibfnamefont {A.}~\bibnamefont {Golnik}}, \bibinfo {author} {\bibfnamefont
  {J.~A.}\ \bibnamefont {Gaj}}, \bibinfo {author} {\bibfnamefont
  {P.}~\bibnamefont {Kossacki}}, \ and\ \bibinfo {author} {\bibfnamefont
  {P.}~\bibnamefont {Wojnar}},\ }\href {\doibase 10.1103/PhysRevB.79.153301}
  {\bibfield  {journal} {\bibinfo  {journal} {Phys. Rev. B}\ }\textbf {\bibinfo
  {volume} {79}},\ \bibinfo {pages} {153301} (\bibinfo {year}
  {2009})}\BibitemShut {NoStop}%
\bibitem [{\citenamefont {Babinski}\ \emph {et~al.}(2005)\citenamefont
  {Babinski}, \citenamefont {Awirothananon}, \citenamefont {Lapointe},
  \citenamefont {Wasilewski}, \citenamefont {Raymond},\ and\ \citenamefont
  {Potemski}}]{adamphysE}%
  \BibitemOpen
  \bibfield  {author} {\bibinfo {author} {\bibfnamefont {A.}~\bibnamefont
  {Babinski}}, \bibinfo {author} {\bibfnamefont {S.}~\bibnamefont
  {Awirothananon}}, \bibinfo {author} {\bibfnamefont {J.}~\bibnamefont
  {Lapointe}}, \bibinfo {author} {\bibfnamefont {Z.}~\bibnamefont
  {Wasilewski}}, \bibinfo {author} {\bibfnamefont {S.}~\bibnamefont {Raymond}},
  \ and\ \bibinfo {author} {\bibfnamefont {M.}~\bibnamefont {Potemski}},\
  }\href {\doibase 10.1016/j.physe.2004.08.050} {\bibfield  {journal} {\bibinfo
   {journal} {Physica E}\ }\textbf {\bibinfo {volume} {26}},\ \bibinfo {pages}
  {190} (\bibinfo {year} {2005})}\BibitemShut {NoStop}%
\bibitem [{\citenamefont {Gammon}\ \emph {et~al.}(1996)\citenamefont {Gammon},
  \citenamefont {Snow}, \citenamefont {Shanabrook}, \citenamefont {Katzer},\
  and\ \citenamefont {Park}}]{gammonprl}%
  \BibitemOpen
  \bibfield  {author} {\bibinfo {author} {\bibfnamefont {D.}~\bibnamefont
  {Gammon}}, \bibinfo {author} {\bibfnamefont {E.~S.}\ \bibnamefont {Snow}},
  \bibinfo {author} {\bibfnamefont {B.~V.}\ \bibnamefont {Shanabrook}},
  \bibinfo {author} {\bibfnamefont {D.~S.}\ \bibnamefont {Katzer}}, \ and\
  \bibinfo {author} {\bibfnamefont {D.}~\bibnamefont {Park}},\ }\href {\doibase
  10.1103/PhysRevLett.76.3005} {\bibfield  {journal} {\bibinfo  {journal}
  {Phys. Rev. Lett.}\ }\textbf {\bibinfo {volume} {76}},\ \bibinfo {pages}
  {3005} (\bibinfo {year} {1996})}\BibitemShut {NoStop}%
\bibitem [{\citenamefont {Favero}\ \emph {et~al.}(2005)\citenamefont {Favero},
  \citenamefont {Cassabois}, \citenamefont {Voisin}, \citenamefont {Delalande},
  \citenamefont {Roussignol}, \citenamefont {Ferreira}, \citenamefont
  {Couteau}, \citenamefont {Poizat},\ and\ \citenamefont {G\'erard}}]{favero}%
  \BibitemOpen
  \bibfield  {author} {\bibinfo {author} {\bibfnamefont {I.}~\bibnamefont
  {Favero}}, \bibinfo {author} {\bibfnamefont {G.}~\bibnamefont {Cassabois}},
  \bibinfo {author} {\bibfnamefont {C.}~\bibnamefont {Voisin}}, \bibinfo
  {author} {\bibfnamefont {C.}~\bibnamefont {Delalande}}, \bibinfo {author}
  {\bibfnamefont {P.}~\bibnamefont {Roussignol}}, \bibinfo {author}
  {\bibfnamefont {R.}~\bibnamefont {Ferreira}}, \bibinfo {author}
  {\bibfnamefont {C.}~\bibnamefont {Couteau}}, \bibinfo {author} {\bibfnamefont
  {J.~P.}\ \bibnamefont {Poizat}}, \ and\ \bibinfo {author} {\bibfnamefont
  {J.~M.}\ \bibnamefont {G\'erard}},\ }\href {\doibase
  10.1103/PhysRevB.71.233304} {\bibfield  {journal} {\bibinfo  {journal} {Phys.
  Rev. B}\ }\textbf {\bibinfo {volume} {71}},\ \bibinfo {pages} {233304}
  (\bibinfo {year} {2005})}\BibitemShut {NoStop}%
\bibitem [{\citenamefont {Flissikowski}\ \emph {et~al.}(2001)\citenamefont
  {Flissikowski}, \citenamefont {Hundt}, \citenamefont {Lowisch}, \citenamefont
  {Rabe},\ and\ \citenamefont {Henneberger}}]{flissikowski}%
  \BibitemOpen
  \bibfield  {author} {\bibinfo {author} {\bibfnamefont {T.}~\bibnamefont
  {Flissikowski}}, \bibinfo {author} {\bibfnamefont {A.}~\bibnamefont {Hundt}},
  \bibinfo {author} {\bibfnamefont {M.}~\bibnamefont {Lowisch}}, \bibinfo
  {author} {\bibfnamefont {M.}~\bibnamefont {Rabe}}, \ and\ \bibinfo {author}
  {\bibfnamefont {F.}~\bibnamefont {Henneberger}},\ }\href {\doibase
  10.1103/PhysRevLett.86.3172} {\bibfield  {journal} {\bibinfo  {journal}
  {Phys. Rev. Lett.}\ }\textbf {\bibinfo {volume} {86}},\ \bibinfo {pages}
  {3172} (\bibinfo {year} {2001})}\BibitemShut {NoStop}%
\bibitem [{\citenamefont {Bayer}\ \emph {et~al.}(2002)\citenamefont {Bayer},
  \citenamefont {Ortner}, \citenamefont {Stern}, \citenamefont {Kuther},
  \citenamefont {Gorbunov}, \citenamefont {Forchel}, \citenamefont {Hawrylak},
  \citenamefont {Fafard}, \citenamefont {Hinzer}, \citenamefont {Reinecke},
  \citenamefont {Walck}, \citenamefont {Reithmaier}, \citenamefont {Klopf},\
  and\ \citenamefont {Schafer}}]{bayerprb}%
  \BibitemOpen
  \bibfield  {author} {\bibinfo {author} {\bibfnamefont {M.}~\bibnamefont
  {Bayer}}, \bibinfo {author} {\bibfnamefont {G.}~\bibnamefont {Ortner}},
  \bibinfo {author} {\bibfnamefont {O.}~\bibnamefont {Stern}}, \bibinfo
  {author} {\bibfnamefont {A.}~\bibnamefont {Kuther}}, \bibinfo {author}
  {\bibfnamefont {A.~A.}\ \bibnamefont {Gorbunov}}, \bibinfo {author}
  {\bibfnamefont {A.}~\bibnamefont {Forchel}}, \bibinfo {author} {\bibfnamefont
  {P.}~\bibnamefont {Hawrylak}}, \bibinfo {author} {\bibfnamefont
  {S.}~\bibnamefont {Fafard}}, \bibinfo {author} {\bibfnamefont
  {K.}~\bibnamefont {Hinzer}}, \bibinfo {author} {\bibfnamefont {T.~L.}\
  \bibnamefont {Reinecke}}, \bibinfo {author} {\bibfnamefont {S.~N.}\
  \bibnamefont {Walck}}, \bibinfo {author} {\bibfnamefont {J.~P.}\ \bibnamefont
  {Reithmaier}}, \bibinfo {author} {\bibfnamefont {F.}~\bibnamefont {Klopf}}, \
  and\ \bibinfo {author} {\bibfnamefont {F.}~\bibnamefont {Schafer}},\ }\href
  {\doibase 10.1103/PhysRevB.65.195315} {\bibfield  {journal} {\bibinfo
  {journal} {Phys. Rev. B}\ }\textbf {\bibinfo {volume} {65}},\ \bibinfo
  {pages} {195315} (\bibinfo {year} {2002})}\BibitemShut {NoStop}%
\bibitem [{\citenamefont {Karlsson}\ \emph {et~al.}(2010)\citenamefont
  {Karlsson}, \citenamefont {Dupertuis}, \citenamefont {Oberli}, \citenamefont
  {Pelucchi}, \citenamefont {Rudra}, \citenamefont {Holtz},\ and\ \citenamefont
  {Kapon}}]{karlsson}%
  \BibitemOpen
  \bibfield  {author} {\bibinfo {author} {\bibfnamefont {K.~F.}\ \bibnamefont
  {Karlsson}}, \bibinfo {author} {\bibfnamefont {M.~A.}\ \bibnamefont
  {Dupertuis}}, \bibinfo {author} {\bibfnamefont {D.~Y.}\ \bibnamefont
  {Oberli}}, \bibinfo {author} {\bibfnamefont {E.}~\bibnamefont {Pelucchi}},
  \bibinfo {author} {\bibfnamefont {A.}~\bibnamefont {Rudra}}, \bibinfo
  {author} {\bibfnamefont {P.~O.}\ \bibnamefont {Holtz}}, \ and\ \bibinfo
  {author} {\bibfnamefont {E.}~\bibnamefont {Kapon}},\ }\href {\doibase
  10.1103/PhysRevB.81.161307} {\bibfield  {journal} {\bibinfo  {journal} {Phys.
  Rev. B}\ }\textbf {\bibinfo {volume} {81}},\ \bibinfo {pages} {161307}
  (\bibinfo {year} {2010})}\BibitemShut {NoStop}%
\bibitem [{\citenamefont {Molas}\ \emph {et~al.}(2013)\citenamefont {Molas},
  \citenamefont {Nicolet}, \citenamefont {Potemski},\ and\ \citenamefont
  {Babi{\'n}ski}}]{molascharged}%
  \BibitemOpen
  \bibfield  {author} {\bibinfo {author} {\bibfnamefont {M.~R.}\ \bibnamefont
  {Molas}}, \bibinfo {author} {\bibfnamefont {A.~A.~L.}\ \bibnamefont
  {Nicolet}}, \bibinfo {author} {\bibfnamefont {M.}~\bibnamefont {Potemski}}, \
  and\ \bibinfo {author} {\bibfnamefont {A.}~\bibnamefont {Babi{\'n}ski}},\
  }\href {\doibase 10.12693/APhysPolA.124.785} {\bibfield  {journal} {\bibinfo
  {journal} {Acta Phys. Pol. A}\ }\textbf {\bibinfo {volume} {124}},\ \bibinfo
  {pages} {785} (\bibinfo {year} {2013})}\BibitemShut {NoStop}%
\bibitem [{\citenamefont {Warming}\ \emph {et~al.}(2009)\citenamefont
  {Warming}, \citenamefont {Siebert}, \citenamefont {Schliwa}, \citenamefont
  {Stock}, \citenamefont {Zimmermann},\ and\ \citenamefont
  {Bimberg}}]{warming}%
  \BibitemOpen
  \bibfield  {author} {\bibinfo {author} {\bibfnamefont {T.}~\bibnamefont
  {Warming}}, \bibinfo {author} {\bibfnamefont {E.}~\bibnamefont {Siebert}},
  \bibinfo {author} {\bibfnamefont {A.}~\bibnamefont {Schliwa}}, \bibinfo
  {author} {\bibfnamefont {E.}~\bibnamefont {Stock}}, \bibinfo {author}
  {\bibfnamefont {R.}~\bibnamefont {Zimmermann}}, \ and\ \bibinfo {author}
  {\bibfnamefont {D.}~\bibnamefont {Bimberg}},\ }\href {\doibase
  10.1103/PhysRevB.79.125316} {\bibfield  {journal} {\bibinfo  {journal} {Phys.
  Rev. B}\ }\textbf {\bibinfo {volume} {79}},\ \bibinfo {pages} {125316}
  (\bibinfo {year} {2009})}\BibitemShut {NoStop}%
\bibitem [{\citenamefont {Babi{\'n}ski}\ \emph {et~al.}(2006)\citenamefont
  {Babi{\'n}ski}, \citenamefont {Potemski}, \citenamefont {Raymond},
  \citenamefont {Lapointe},\ and\ \citenamefont
  {Wasilewski}}]{babinskiprb2006b}%
  \BibitemOpen
  \bibfield  {author} {\bibinfo {author} {\bibfnamefont {A.}~\bibnamefont
  {Babi{\'n}ski}}, \bibinfo {author} {\bibfnamefont {M.}~\bibnamefont
  {Potemski}}, \bibinfo {author} {\bibfnamefont {S.}~\bibnamefont {Raymond}},
  \bibinfo {author} {\bibfnamefont {J.}~\bibnamefont {Lapointe}}, \ and\
  \bibinfo {author} {\bibfnamefont {Z.~R.}\ \bibnamefont {Wasilewski}},\ }\href
  {\doibase 10.1103/PhysRevB.74.155301} {\bibfield  {journal} {\bibinfo
  {journal} {Phys. Rev. B}\ }\textbf {\bibinfo {volume} {74}},\ \bibinfo
  {pages} {155301} (\bibinfo {year} {2006})}\BibitemShut {NoStop}%
\bibitem [{\citenamefont {Fock}(1928)}]{fock}%
  \BibitemOpen
  \bibfield  {author} {\bibinfo {author} {\bibfnamefont {V.}~\bibnamefont
  {Fock}},\ }\href {\doibase 10.1007/BF01390750} {\bibfield  {journal}
  {\bibinfo  {journal} {Z. Phys.}\ }\textbf {\bibinfo {volume} {47}},\ \bibinfo
  {pages} {446} (\bibinfo {year} {1928})}\BibitemShut {NoStop}%
\bibitem [{\citenamefont {Darwin}(1930)}]{darwin}%
  \BibitemOpen
  \bibfield  {author} {\bibinfo {author} {\bibfnamefont {C.~G.}\ \bibnamefont
  {Darwin}},\ }\href {\doibase 10.1017/S0305004100009373} {\bibfield  {journal}
  {\bibinfo  {journal} {Proc. Cambridge Philos. Soc.}\ }\textbf {\bibinfo
  {volume} {27}},\ \bibinfo {pages} {86} (\bibinfo {year} {1930})}\BibitemShut
  {NoStop}%
\bibitem [{\citenamefont {Raymond}\ \emph {et~al.}(2004)\citenamefont
  {Raymond}, \citenamefont {Studenikin}, \citenamefont {Sachrajda},
  \citenamefont {Wasilewski}, \citenamefont {Cheng}, \citenamefont {Sheng},
  \citenamefont {Hawrylak}, \citenamefont {Babi\'nski}, \citenamefont
  {Potemski}, \citenamefont {Ortner},\ and\ \citenamefont
  {Bayer}}]{raymondprl}%
  \BibitemOpen
  \bibfield  {author} {\bibinfo {author} {\bibfnamefont {S.}~\bibnamefont
  {Raymond}}, \bibinfo {author} {\bibfnamefont {S.}~\bibnamefont {Studenikin}},
  \bibinfo {author} {\bibfnamefont {A.}~\bibnamefont {Sachrajda}}, \bibinfo
  {author} {\bibfnamefont {Z.}~\bibnamefont {Wasilewski}}, \bibinfo {author}
  {\bibfnamefont {S.~J.}\ \bibnamefont {Cheng}}, \bibinfo {author}
  {\bibfnamefont {W.}~\bibnamefont {Sheng}}, \bibinfo {author} {\bibfnamefont
  {P.}~\bibnamefont {Hawrylak}}, \bibinfo {author} {\bibfnamefont
  {A.}~\bibnamefont {Babi\'nski}}, \bibinfo {author} {\bibfnamefont
  {M.}~\bibnamefont {Potemski}}, \bibinfo {author} {\bibfnamefont
  {G.}~\bibnamefont {Ortner}}, \ and\ \bibinfo {author} {\bibfnamefont
  {M.}~\bibnamefont {Bayer}},\ }\href {\doibase 10.1103/PhysRevLett.92.187402}
  {\bibfield  {journal} {\bibinfo  {journal} {Phys. Rev. Lett.}\ }\textbf
  {\bibinfo {volume} {92}},\ \bibinfo {pages} {187402} (\bibinfo {year}
  {2004})}\BibitemShut {NoStop}%
\bibitem [{\citenamefont {Benny}\ \emph {et~al.}(2011)\citenamefont {Benny},
  \citenamefont {Kodriano}, \citenamefont {Poem}, \citenamefont {Khatsevitch},
  \citenamefont {Gershoni},\ and\ \citenamefont {Petroff}}]{benny2011}%
  \BibitemOpen
  \bibfield  {author} {\bibinfo {author} {\bibfnamefont {Y.}~\bibnamefont
  {Benny}}, \bibinfo {author} {\bibfnamefont {Y.}~\bibnamefont {Kodriano}},
  \bibinfo {author} {\bibfnamefont {E.}~\bibnamefont {Poem}}, \bibinfo {author}
  {\bibfnamefont {S.}~\bibnamefont {Khatsevitch}}, \bibinfo {author}
  {\bibfnamefont {D.}~\bibnamefont {Gershoni}}, \ and\ \bibinfo {author}
  {\bibfnamefont {P.~M.}\ \bibnamefont {Petroff}},\ }\href {\doibase
  10.1103/PhysRevB.84.075473} {\bibfield  {journal} {\bibinfo  {journal} {Phys.
  Rev. B}\ }\textbf {\bibinfo {volume} {84}},\ \bibinfo {pages} {075473}
  (\bibinfo {year} {2011})}\BibitemShut {NoStop}%
\bibitem [{\citenamefont {Molas}\ \emph {et~al.}(2014)\citenamefont {Molas},
  \citenamefont {Nicolet}, \citenamefont {Pi\k{e}tka}, \citenamefont
  {Babi\'nski},\ and\ \citenamefont {Potemski}}]{molasple}%
  \BibitemOpen
  \bibfield  {author} {\bibinfo {author} {\bibfnamefont {M.~R.}\ \bibnamefont
  {Molas}}, \bibinfo {author} {\bibfnamefont {A.~A.~L.}\ \bibnamefont
  {Nicolet}}, \bibinfo {author} {\bibfnamefont {B.}~\bibnamefont {Pi\k{e}tka}},
  \bibinfo {author} {\bibfnamefont {A.}~\bibnamefont {Babi\'nski}}, \ and\
  \bibinfo {author} {\bibfnamefont {M.}~\bibnamefont {Potemski}},\ }\href
  {\doibase 10.12693/APhysPolA.126.1066} {\bibfield  {journal} {\bibinfo
  {journal} {Acta Phys. Pol. A}\ }\textbf {\bibinfo {volume} {126}},\ \bibinfo
  {pages} {1066} (\bibinfo {year} {2014})}\BibitemShut {NoStop}%
\bibitem [{\citenamefont {Molas}\ \emph {et~al.}()\citenamefont {Molas},
  \citenamefont {Nicolet}, \citenamefont {Pi\k{e}tka}, \citenamefont
  {Babi{\'n}ski},\ and\ \citenamefont {Potemski}}]{molasX+}%
  \BibitemOpen
  \bibfield  {author} {\bibinfo {author} {\bibfnamefont {M.~R.}\ \bibnamefont
  {Molas}}, \bibinfo {author} {\bibfnamefont {A.~A.~L.}\ \bibnamefont
  {Nicolet}}, \bibinfo {author} {\bibfnamefont {B.}~\bibnamefont {Pi\k{e}tka}},
  \bibinfo {author} {\bibfnamefont {A.}~\bibnamefont {Babi{\'n}ski}}, \ and\
  \bibinfo {author} {\bibfnamefont {M.}~\bibnamefont {Potemski}},\ }\href@noop
  {} {\ }\Eprint {http://arxiv.org/abs/1602.03789} {arXiv:1602.03789}
  \BibitemShut {NoStop}%
\end{thebibliography}%

\end{document}